\documentclass[jkps,preprint,fleqn,showpacs,showkeys]{revtex4}
\usepackage{graphicx}
\usepackage{amssymb}
\usepackage{amsmath}
\usepackage{bm}
\usepackage{hyperref}

\def\la{\langle}
\def\ra{\rangle}

\begin{document}

\setcounter{page}{0}
\title[]{Explicit flavor symmetry breaking and holographic compact stars}
\author{Youngman \surname{Kim}}
\email{ykim@ibs.re.kr}
\affiliation{Rare Isotope Science Project, Institute for Basic Science, Daejeon 305-811, Korea}
\author{Chang-Hwan \surname{Lee}}
\email{clee@pusan.ac.kr}
\thanks{Visiting Scholar in Department of Physics and Astronomy, State University of New York, Stony Brook, NY 11794, USA, during Aug. 2013 $\sim$ Aug. 2014}
\affiliation{Department of Physics, Pusan National University, Busan 609-735, Korea}
\author{Ik Jae \surname{Shin}}
\email{geniean@ibs.re.kr}
\affiliation{Rare Isotope Science Project, Institute for Basic Science, Daejeon 305-811, Korea}
\author{Mew-Bing \surname{Wan}}
\email{mbwan@apctp.org}
\affiliation{Asia-Pacific Center for Theoretical Physics, Pohang, Gyeongbuk 790-784, Korea}

\date[]{}

\begin{abstract}
We study the effects of flavor symmetry breaking on holographic dense matter and compact stars in the D4/D6 model. To this end, two light  flavors and one intermediate mass flavor are considered.
For two light quarks, we investigate how the strong isospin violation affects the properties of holographic dense matter and compact stars.
We observe that quark-antiquark condensates  are flavor dependent and show interesting behavior near the transition from dense matter with only one flavor to matter with two flavors.
An  intermediate mass quark is introduced to investigate the role of  the third flavor. The mass-radius relations of holographic compact stars with three flavors show that the mass-radius curve changes drastically at a transition density from which the third flavor begins to appear in the matter.
\end{abstract}



\maketitle

\flushbottom

\section{Introduction}

Based on the Anti de Sitter/conformal field theory (AdS/CFT) correspondence \cite{Maldacena:1997re,Gubser:1998bc,Witten:1998qj},  possibilities of studying strongly interacting systems such as quantum chromodynamics (QCD) at low energy have been explored.
There are in general two distinct ways to construct holographic dual models of strongly interacting systems. One
is a top-down approach based on stringy D-brane configurations. Another is a bottom-up approach, where a 5D
holographic dual is constructed from a boundary field theory such as QCD. For reviews on holographic QCD we refer to \cite{review_hQCD}.

Compact stars that we are interested in are neutron stars in which strong interaction dominates. Neutron stars have been observed mainly in radio pulsars and low mass X-ray binaries (LMXB). Most of the radio pulsars are single neutron stars with which mass measurement is very hard, hence neutron star masses have been estimated mainly in binaries \cite{Lattimer:2006xb,Prakash:2013fha}. The maximum mass of neutron star is believed to be larger than $2 M_\odot$, the mass of neutron stars in  neutron star-white dwarf binaries J1614$-$2230 \cite{Demorest:2010bx}
and J0348$+$0432 \cite{Antoniadis:2013pzd}.
In addition, possibilities of estimating both mass and radius of neutron star in LMXB have been suggested
\cite{Lattimer:2013hma,Guver:2011qw,Guver:2011js}.
Based on these observations, many neutron star equations of state have been ruled out. However, due to the uncertainties in the equations of state at high densities beyond normal nuclear matter density, the inner structure of neutron star is still unknown.

Symmetry energy and the strange quark may play a very important role in describing the central region of neutron stars. Symmetry energy is the energy difference due to the asymmetry in isospin, i.e., the number of protons and neutrons. After the discovery of $2 M_\odot$ neutron stars \cite{Demorest:2010bx,Antoniadis:2013pzd} harder equations of states are favored in many literature and the symmetry energy turned out to be  crucial in making the harder equations of states. On the other hand, as density increases, the Fermi energy of up and down quarks (neutrons and protons) becomes very high and the  strange quark (hyperon) should come in because the introduction of one more quark degree of freedom can reduce the total energy of the system. The strange quark is neither light nor heavy compared to the other quarks in the density region which we are interested in. In many effective models, the strange quark has been treated separately from up and down quarks mainly due to the current quark mass differences.


In our previous work \cite{Kim:2011da}, we used the equation of state (EoS) from holographic QCD models to see if any holographic QCD models
support a stable compact star. Even though the radii of holographic stars are much larger than those of observed neutron stars, we found that the stable compact stars are possible in the D4/D6 model. For a recent study on holographic neutron stars based on the probe D8 brane we refer to \cite{Ghoroku:2013gja}.
In the present work, we use the D4/D6 model (in the confining phase) to investigate the effects of flavor symmetry breaking on holographic dense matter and compact stars.
As well known, one of the excellent points of the D4/D6 model is that the model can easily incorporate non-zero quark masses.
Therefore, the D4/D6 model
may be suitable for investigating flavor physics, especially phenomena associated with explicit flavor symmetry breaking due to the quark mass differences.

In Sec.~\ref{sec:D4D6} we introduce the model  and  discuss previous works in the D4/D6 model with different quark masses.
In Sec.~\ref{sec:split} we discuss the meson mass splitting and demonstrate that the model might be  an adequate tool for investigating asymmetric dense matter with ${N_f>1}$.
In Sec.~\ref{sec:cond} the quark-antiquark condensates are discussed to see their flavor dependence in the asymmetric dense matter.
 In Sec.~\ref{sec:massdep} we study the effects of explicit flavor symmetry breaking on the properties of compact stars.
Our discussion follows in Sec.~\ref{sec:sum}.

\section{D4/D6 model with explicit flavor symmetry breaking}
\label{sec:D4D6}

In this section, we introduce the D4/D6 model \cite{Kruczenski:2003uq} and its extended version in (asymmetric) dense matter with $N_f\ge 1$ \cite{Seo:2008qc,Kim:2009ey, Kim:2011gw}.
In the probe approximation, the $N_c$ D4 branes are replaced by the corresponding type IIA SUGRA background while $N_f$ D6 branes are treated as probes.
The solution of the confining D4 brane reads
\begin{eqnarray}
\label{backgroundD6}
ds^2&\!=\!&\Big(\frac{U}{L}\Big)^{3/2}\big(\eta_{\mu\nu}dx^\mu dx^\nu+f(U)d\tau^2\big)+\Big(\frac{L}{U}\Big)^{3/2}\Big(\frac{dU^2}{f(U)}+U^2d\Omega_4^2\Big) \nonumber \\
	&\!=\!&\Big(\frac{U}{L}\Big)^{3/2}\big(\eta_{\mu\nu}dx^\mu dx^\nu+f(U)d\tau^2\big)+\Big(\frac{L}{U}\Big)^{3/2}\Big(\frac{U}{\xi}\Big)^2(d\xi^2+\xi^2d\Omega_4^2)\,, \\
	&&~~~~~~~~~~e^\phi=g_s\Big(\frac{U}{L}\Big)^{3/4}~~~,~~~~F_4=dC_3=\frac{2\pi N_c\epsilon_4}{\Omega_4}\,,
\end{eqnarray}
where $f(U)=1-(U_\textrm{KK}/U)^3\,$ and $\epsilon_4$ and $\Omega_4$ denote the volume form and volume of the unit $S^4$, respectively.
The D4 branes are compactified in the $\tau$-direction, and $U$ is considered as a radial coordinate of the transverse to the world-volume of branes.
The dimensionless radial coordinate $\xi$ is defined by
\begin{equation}
	\Big(\frac{U}{U_\textrm{KK}}\Big)^{3/2}\equiv\frac{1}{2}\big(\xi^{3/2}+\frac{1}{\xi^{3/2}}\big)\,.
 \end{equation}
 The parameters in string theory are related to those in  gauge theory as
 \begin{equation}
	L^3=\frac{\lambda l_s^2}{2M_\textrm{KK}}~~,~~~~U_\textrm{KK}=\frac{2\lambda M_\textrm{KK}l_s^2}{9}~~,~~~~g_s=\frac{\lambda}{2\pi N_c M_\textrm{KK} l_s}
 \end{equation}
 with the 't$\,$Hooft coupling $\lambda\equiv g_\textrm{YM}^2 N_c$.
 To describe the D6 brane configuration, the background metric Eq.~\eqref{backgroundD6} is rewritten as
 \begin{equation}
	ds^2=\Big(\frac{U}{L}\Big)^{3/2}\big(\eta_{\mu\nu}dx^\mu dx^\nu+f(U)d\tau^2\big)+\Big(\frac{L}{U}\Big)^{3/2}\Big(\frac{U}{\xi}\Big)^2\big(dz^2+z^2d\Omega^2_2+dr^2+r^2d\phi^2\big)\, ,
 \end{equation}
where $\xi^2=z^2+r^2$. The distance between D4 and D6 branes at $z=\infty$ is interpreted as quark mass and the coefficient of the second term corresponds to the quark-antiquark condensate; $r(z\rightarrow \infty)\sim m_q +c_q/z\,$. Then, the  Dirac-Born-Infeld (DBI) action for the D6 brane, in the case of $N_f=1$, becomes
\begin{eqnarray}
\label{DBIactionD6}
	S_\textrm{D6}&\!=\!&-\mu_6\!\int\!d^7\sigma~e^{-\phi}\!\sqrt{-\textrm{det}(\textrm{P}[g]+2\pi\alpha^\prime F)} \nonumber \\
	&\!=\!&-\tau_6\int dt dz\;z^2(1+1/\xi^3)^{4/3}\sqrt{(1+1/\xi^3)^{4/3}(1+{r^\prime}^2)-\tilde{F}^2}
 \end{eqnarray}
 where $\mu_6^{-1}=(2\pi)^6 l_s^7\;,~\tau_6=\mu_6 g_s^{-1}\Omega_2 V_3(U_\textrm{KK}^3/4)$ and $\tilde{F}=(2^{2/3}/U_\textrm{KK})2\pi\alpha^\prime F_{zt}=\tilde{A}_t^\prime\,$.
The source of the gauge field is the end point of the fundamental strings.
We define the dimensionless quantity $\tilde{Q}$,
\begin{equation}
	\frac{\partial\mathcal{L}_\textrm{D6}}
{\partial\tilde{F}}=\frac{z^2(1+1/\xi^3)^{4/3}\tilde{F}}{\sqrt{(1+1/\xi^3)^{4/3}(1+{r^\prime}^2)-\tilde{F}^2}}\equiv\tilde{Q}\, ,
\end{equation}
where  $\tilde{Q}$ is related to the number of fundamental strings $Q$ through $\tilde{Q}=Q/(2\pi\alpha^\prime\tau_6)(U_\textrm{KK}/2^{2/3})$.
The Hamiltonian of the D6 brane is obtained through the Legendre transformation,
\begin{equation} \label{HamiltonianD6}
H_\textrm{D6}=\tau_6\int dz\;\sqrt{(1+1/\xi^3)^{4/3}\left[z^4(1+1/\xi^3)^{8/3}+\tilde{Q}^2\right](1+{r^\prime}^2)}\, .
\end{equation}
This D4/D6 model is extended to dense matter in the confined phase for $N_f=1$ \cite{Seo:2008qc} and for $N_f=2$ \cite{Kim:2009ey}.
(For details on the D4/D6 model in dense matter, we refer to \cite{Seo:2008qc, Kim:2009ey, Witten:1998xy}.)
A compact D4 brane wrapping on the 4-sphere $S^4$ transverse to $\mathbb{R}^{1,3}$ is interpreted as a baryon \cite{Witten:1998xy}.
To describe the compact D4 brane, the $S^4$ part of the background metric  is rewritten as $d\Omega_4^2=d\theta^2+\sin^2\!\theta d\Omega_3^2\,$.
The world-volume coordinates of the compact D4 is $(\,t, \theta, \varphi_\alpha)$, where $\theta$ is the polar angle from the south pole. To study dense matter, we turn on the time component of the gauge field $A_t$.
The $SO(4)$ symmetry gives the ansatz that all fields depend on $\theta$, i.e., $\xi(\theta)$ and $A_t(\theta)$.
The induced metric on this D4 brane is
 \begin{equation} \label{inducedD4}
	ds_\textrm{D4}^2=-\Big(\frac{U}{L}\Big)^{3/2}dt^2+(L^3 U)^{1/2}\left[\left(1+\frac{\dot{\xi}^2}{\xi^2}\right)d\theta^2+\sin^2\!\theta d\Omega_3^2\right]\, ,
 \end{equation}
where $\dot{\xi}=\partial\xi/\partial\theta$.
 The DBI action of the compact D4 brane is then
 \begin{eqnarray} \label{DBIactionD4}
	S_\textrm{D4}=\int\!dt d\theta\,\sin^3\!\theta\sqrt{(1+1/\xi^3)^{4/3}(\xi^2+\dot{\xi}^2)-\tilde{F}^2}\,,
 \end{eqnarray}
 where $\mu_4^{-1}=(2\pi)^4 l_s^5\;,~\tau_4=\mu_4 g_s^{-1}\Omega_3 L^3(U_\textrm{KK}/2^{2/3})$ and $\tilde{F}=(2^{2/3}/U_\textrm{KK})2\pi\alpha^\prime F_{\theta t}=\dot{\tilde{A}}_t\,$.
The Hamiltonian of the compact D4 brane is also given by the Legendre transformation,
\begin{equation} \label{HamiltonianD4}
  	H_\textrm{D4}=\tau_4\int d\theta\;\sqrt{(1+1/\xi^3)^{4/3}(\xi^2+\dot{\xi}^2)(\tilde{D}^2+\sin^6\theta)}\, ,
 \end{equation}
 where $\tilde{D}$ is given by
 \begin{equation}
	\tilde{D}(\theta)=2-(3\cos\theta-\cos^3\theta)\, .
 \end{equation}

Now, we consider a stability of the D4/D6 system in dense matter.
Since $N_c$ fundamental strings are attached on each compact D4 brane, there exist $N_B(=Q/N_c)$ number of D4 branes.
 The other end of the fundamental string is attached to the D6 brane as a source of the gauge field.
 Since the tension of the string is always larger than that of the brane, the fundamental strings pull both D4 and D6 branes,
 and then the compact D4 and D6 branes meet eventually at a point.
 To secure the stability of the system, the force at the cusp should be balanced, i.e.,
  \begin{equation}
 (Q/N_c)f_\textrm{D4}=f_\textrm{D6}[Q]\, .\label{FBc}
  \end{equation}
  The force due to the brane tension at the cusp is given by the variation of its Hamiltonian with respect to the position of contact point,
 \begin{equation}
	f\equiv\frac{\partial H}{\partial\xi_c}\Big|_\textrm{fixed other values}\, .
 \end{equation}

  At zero temperature, energy density $\epsilon$ and pressure $p$ are calculated as
 \begin{equation}\label{eq:eos}
	 \epsilon=\frac{1}{V_3}\Big(\frac{Q}{N_c}\bar{\Omega}_\textrm{D4}+\bar{\Omega}_\textrm{D6}^\textrm{reg.}[Q]\Big)~,~~
~p=\frac{1}{V_3}\Omega_\textrm{D6}^\textrm{reg.}[Q]\, ,
 \end{equation}
 where $\Omega$ and $\bar{\Omega}$ denote the grand canonical potential and its Legendre transform, respectively.
 These  can be obtained from the on-shell action and Hamiltonian,
 \begin{equation}
	\Omega\equiv S|_\textrm{on-shell}~,~~~~\bar\Omega\equiv H |_\textrm{on-shell}\,.
 \end{equation}
 When we calculate the on-shell values of the probe D6 branes, the integration with the infinite range of $z$ diverges.
 Therefore, we should subtract out the zero-density contribution.
 \begin{equation}
	\Omega^\textrm{reg.}[Q]\equiv\Omega[Q]-\Omega[Q=0]
 \end{equation}

\section{Meson mass splitting in asymmetric dense matter}
\label{sec:split}

In this section, we recapitulate the meson mass splitting done in \cite{Kim:2011gw}
to demonstrate the usefulness of the D4/D6 model in studying flavor physics such as
(strong) isospin symmetry breaking.
To study the dynamics of multi D-branes, one has to start with a non-Abelian DBI action.
We use the equation of motion for vector mesons obtained in \cite{Kim:2011gw} and calculate the properties of  off-diagonal vector mesons 
(for instance, $\rho^\pm$ in nature).
Though we may have to use $m_2/m_1 \simeq 2$ to discuss realistic isospin physics,
for illustration purpose, we take $m_2/m_1=10$ with $\lambda=10$ and $M_\textrm{KK}=1.04$ GeV.

\begin{figure}
\begin{center}
\includegraphics[scale=0.42]{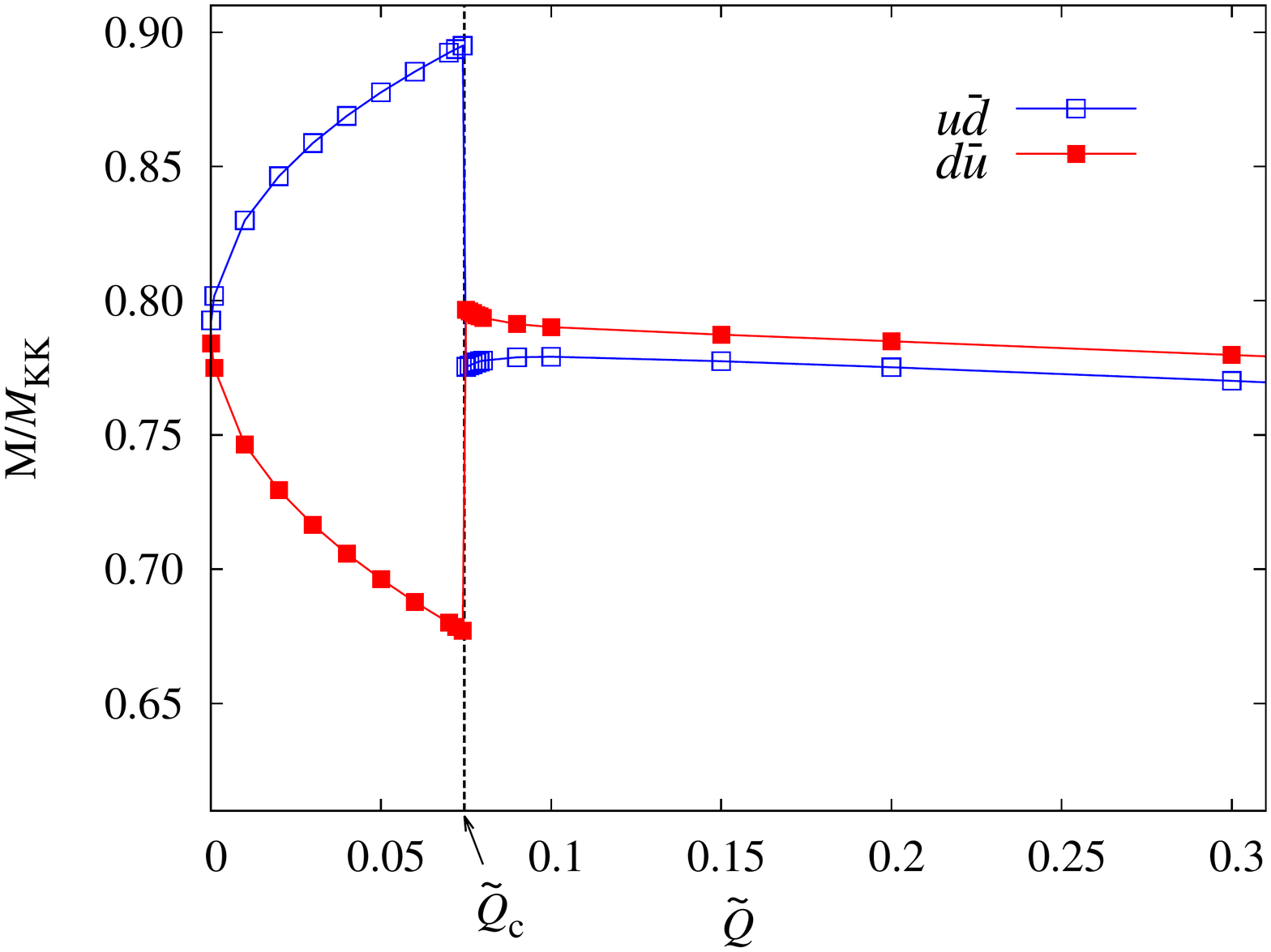}
\caption{An example of meson mass splitting in asymmetric dense matter with $\lambda=10$ and $M_\textrm{KK}=1.04$ GeV. The blue (red) curve is analogous to $u\bar d$ ($d\bar u$).  $\tilde Q_c$ denotes a (dimensionless) transition density from pure $up$ quark matter to matter with both $up$ and $down$ quarks.}
\label{fig:rhoM}
\end{center}
\end{figure}

\begin{figure}
\begin{center}
\includegraphics[scale=0.42]{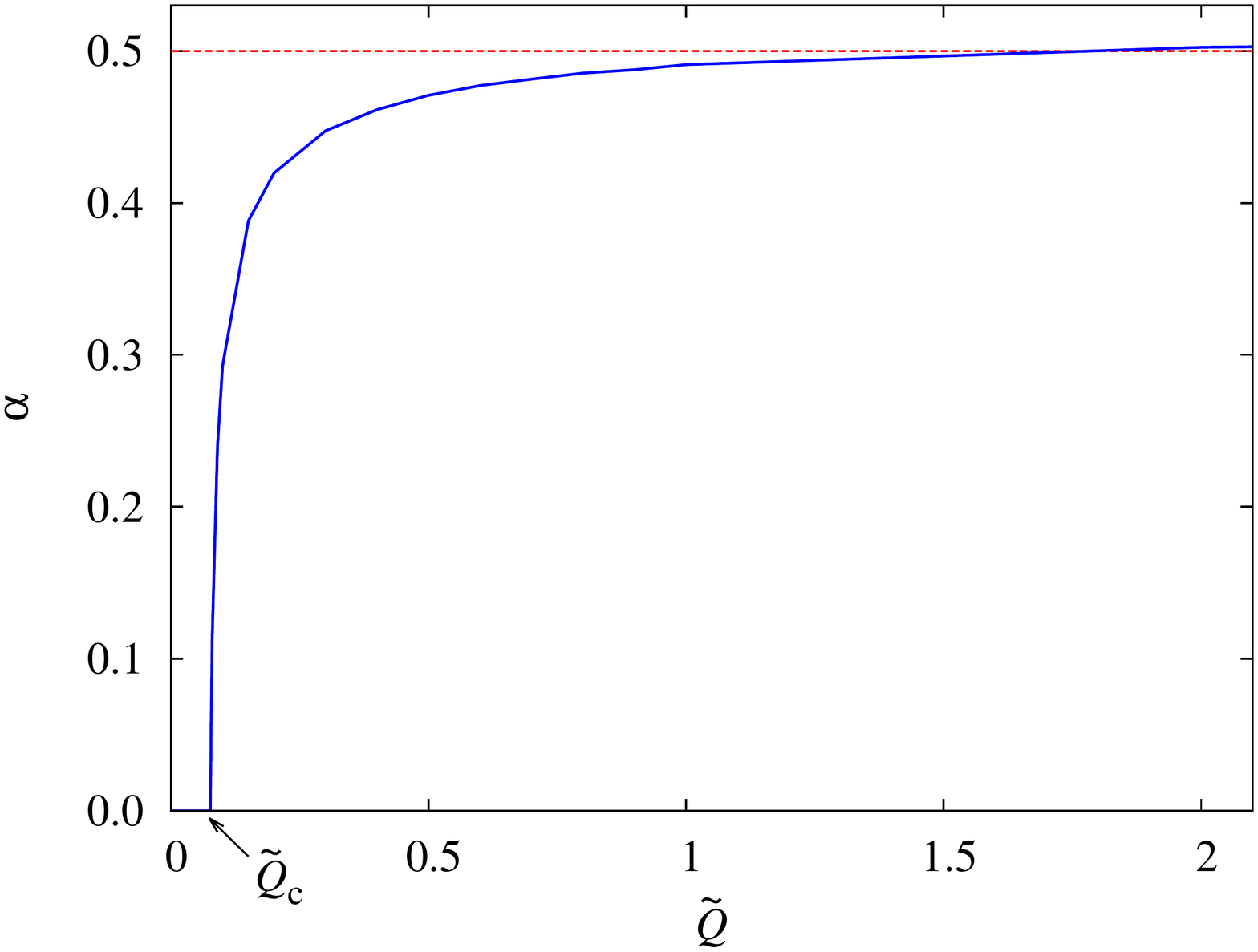}
\caption{The fraction of the $down$ quark in the ground state with $\lambda=10$ and $M_\textrm{KK}=1.04$ GeV. Here $\alpha$ is defined by $\tilde Q_2=\alpha \tilde Q$.}
\label{fig:tQa}
\end{center}
\end{figure}

The result is shown in Fig.~\ref{fig:rhoM}.
Note that similar results were obtained in \cite{Kim:2011gw} with $m_2/m_1=3, 30$.
In this figure the vertical dashed line denotes the transition point from dense matter with one flavor of mass $m_1$
to that with two flavors (analogous to the transition from the pure neutron matter to the symmetric matter).
Before the transition, as expected, mass splitting between two vector mesons with opposite isospins
naturally occurs. After the transition, it shows an interesting  behavior near the transition density.
As discussed in \cite{Kim:2011gw}, we could schematically understand this phenomenon as follows.
Suppose we take quarks with masses $m_1$ and $m_2$ as $up$ and $down$ quarks, respectively.
The off-diagonal mesons, such as $\rho^\pm$, consist of $u\bar d$ ($\rho^+$) and $d \bar u$ ($\rho^-$).
The transition is from pure $up$ quark matter to matter with both $up$ and $down$ quarks.
Due to the Pauli exclusion principle between $down$ quarks in $\rho^-$ and those in the matter,
we can expect that after the transition  the mass of $\rho^-$ would increase, while the mass of $\rho^+$ drops much faster since the population of the $up$ quarks  drops after the transition. The drastic change of the meson masses at the transition point is due to the sudden change of the
number densities at the transition point as shown in Fig.~\ref{fig:tQa}. As discussed in \cite{Kim:2011gw}, this behavior is similar to the transition from nuclear matter to hyperon matter studied within conventional mean field models.

In addition to the work summarized above, there are other studies that may imply the relevance of the D4/D6 type model to
physical phenomena with different quark masses.
In \cite{Kim:2009ey},  a transition from matter with only a single flavor (lighter one)  to
 matter with two flavors (lighter and heavier ones) was studied. It is shown that the transition is almost first order and the number of
 heavier flavor quarks jumps rapidly at the transition density. As discussed in \cite{Kim:2009ey}, this result mimics nuclear matter to hyperon matter transition which may exist in nature, especially
 in the interior of neutron stars.
In \cite{KS3}, the symmetry energy as a function of the dimensionless density with three flavors ($m_1=m_2< m_3$)  was calculated in the D4/D6 model.
It is shown that the slope of the symmetry energy curve suddenly decreases at the transition density
 from matter with two light quarks to matter with two light and one intermediate mass quarks.


\section {Quark condensates}
\label{sec:cond}

\begin{figure} 
\centering
\includegraphics[trim = 60 0 20 0, clip, width=0.52\textwidth]{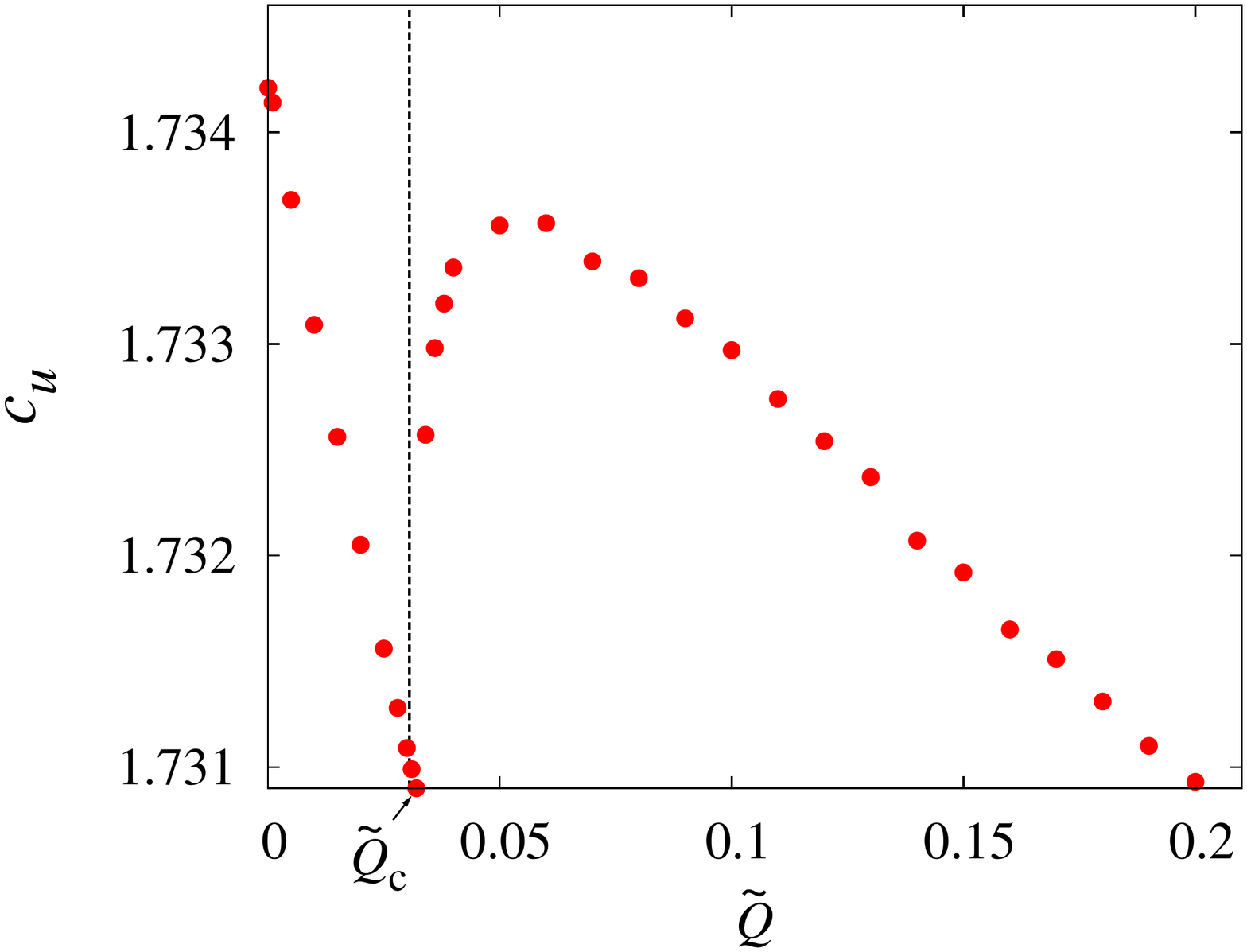}~
\includegraphics[trim = 60 0 20 0, clip, width=0.52\textwidth]{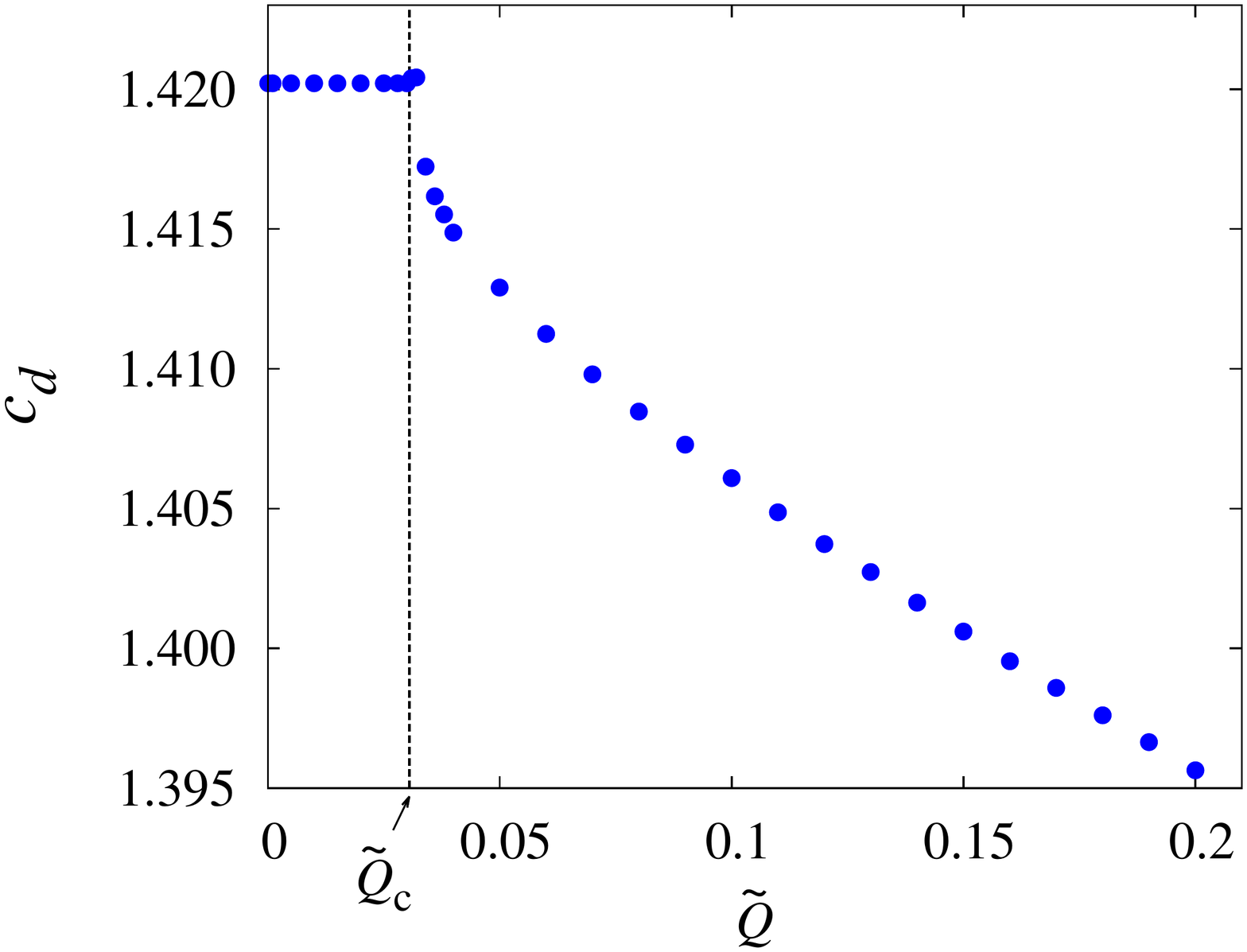}
\caption{Quark-antiquark condensate as a function of $\tilde Q$.
 Here $c_u\sim \la\bar uu\ra$ on the left and  $c_d\sim \la\bar dd\ra$ on the right. }\label{fig:cud}
\end{figure}

In this section we study the effect of explicit flavor symmetry breaking on quark condensates.
We consider a two-flavor system with masses $m_1$ and $m_2$. For simplicity,  we again call a quark with $m_1$ ($m_2$) as $up$ ($down$) quark.
We can easily read off the quark-antiquark condensate from the asymptotic form of the embedding function, $r(z\rightarrow \infty)\sim m_q +c_q/z\,$.
In Fig.~\ref{fig:cud} we show $\la\bar uu\ra$ and $\la \bar dd \ra$   as a function of dimensionless density $\tilde Q$, where $\tilde Q_c$ is  the transition density
from a dense matter with only $up$ quarks in the ground state to a matter with both $up$ and $down$ quarks. Here, we take $m_2/m_1=7$ for illustration purpose. We observed a similar tendency with  $m_2/m_1=3, 15,30$.

In general, one can schematically understand why the quark condensate decreases with density based on the Pauli principle similar to the meson mass splitting.
As the density increases, the low lying phase space relevant for quark-antiquark condensates is occupied by the fermions that constitute the Fermi sea, and thereby forming quark-antiquark condensates requires much energy.
At low densities (before the transition), only $up$ quarks are in the ground state, and   it is expected that  $\la\bar uu\ra$ decreases with density while
$\la \bar dd \ra$ may stay constant.
After the transition, $down$ quarks start to stack in the ground state rapidly as in Fig.~\ref{fig:tQa}  while the density of $up$ quarks is decreasing accordingly. Therefore, it is expected that $\la \bar dd \ra$ drops and  $\la\bar uu\ra$ increases suddenly right after the transition. Our results in Fig.~\ref{fig:cud} near the transition density behave as if there exist a well-defined Fermi sea since they show expected behavior; see, for instance, \cite{Rozali:2007rx} and \cite{Kim:2007vd} for similar observations in a different context.
 But, it is not clear whether the well-defined Fermi sea really exists since both $\la\bar uu\ra$ and $\la \bar dd \ra$ eventually increase with densities at high densities and our basic degrees of freedom are bosons as in the DBI action.

\section{Holographic compact stars: flavor dependence}
\label{sec:massdep}

In this section we study how the quark mass difference and the presence of a third flavor affect the properties of compact stars.
We will focus on the mass and radius of the stars.
For simplicity, we refer to the quarks with masses $m_1$, $m_2$ and $m_3$ as  $up$, $down$ and $strange$ quarks, respectively.


Firstly, we consider a two-flavor system with masses $m_1$ and $m_2$.
As done in \cite{Kim:2009ey} and \cite{Kim:2011gw}, we solve the equations of motion for the D4 and D6 branes numerically using the free parameters $\xi_0$, $\alpha$, $m_1$ and $m_2$,
where $\xi_0$ is the initial value of the position variable of the D4 brane.
We take various mass ratios $m_2/m_1=1,2,7,15,20$ with $m_1=0.1$. Using the relation in \cite{Kim:2010dp}, $m_1=0.1$ corresponds to $23.2$ MeV.
For each $\tilde{Q}$, which represents the total number density of the system, we optimize the parameters $\xi_0$ and $\alpha$ to obtain the lowest energy configuration using the force balance condition between the D4 and D6 branes given in Eq.~\eqref{FBc}.
As shown in \cite{Kim:2009ey} and \cite{Kim:2011gw}, we find that the lowest energy configuration is at a non-zero $\alpha$ starting from a certain transition $\tilde{Q}_c$.
In fact, with these asymmetric mass ratios, i.e., $m_2/m_1 > 1$, the $\alpha$ value increases with increasing density and saturates to $0.5$ as expected.
This way we obtain the energy density and pressure according to Eq.~\eqref{eq:eos} and put them  into the Tolman–Oppenheimer–Volkoff (TOV) equations together with the number density (which is transformed into the matter density) to calculate the mass and radius of a compact star.
The TOV equation is given by
\begin{eqnarray}
\label{eq:tov}
\frac{dp}{dr} = -\frac{1}{2}(\epsilon+p)\frac{2m+8\pi r^3 p}{r(r-2m)},
\end{eqnarray}
where $r$ is the circumferential radius and $\rho$ is the baryonic number density.
The integrated mass of the compact star is obtained through $m(r) \equiv 4\pi\int^r_0\epsilon(r')r'^2 dr'$.

\begin{figure}
\begin{center}
\includegraphics[scale=0.5]{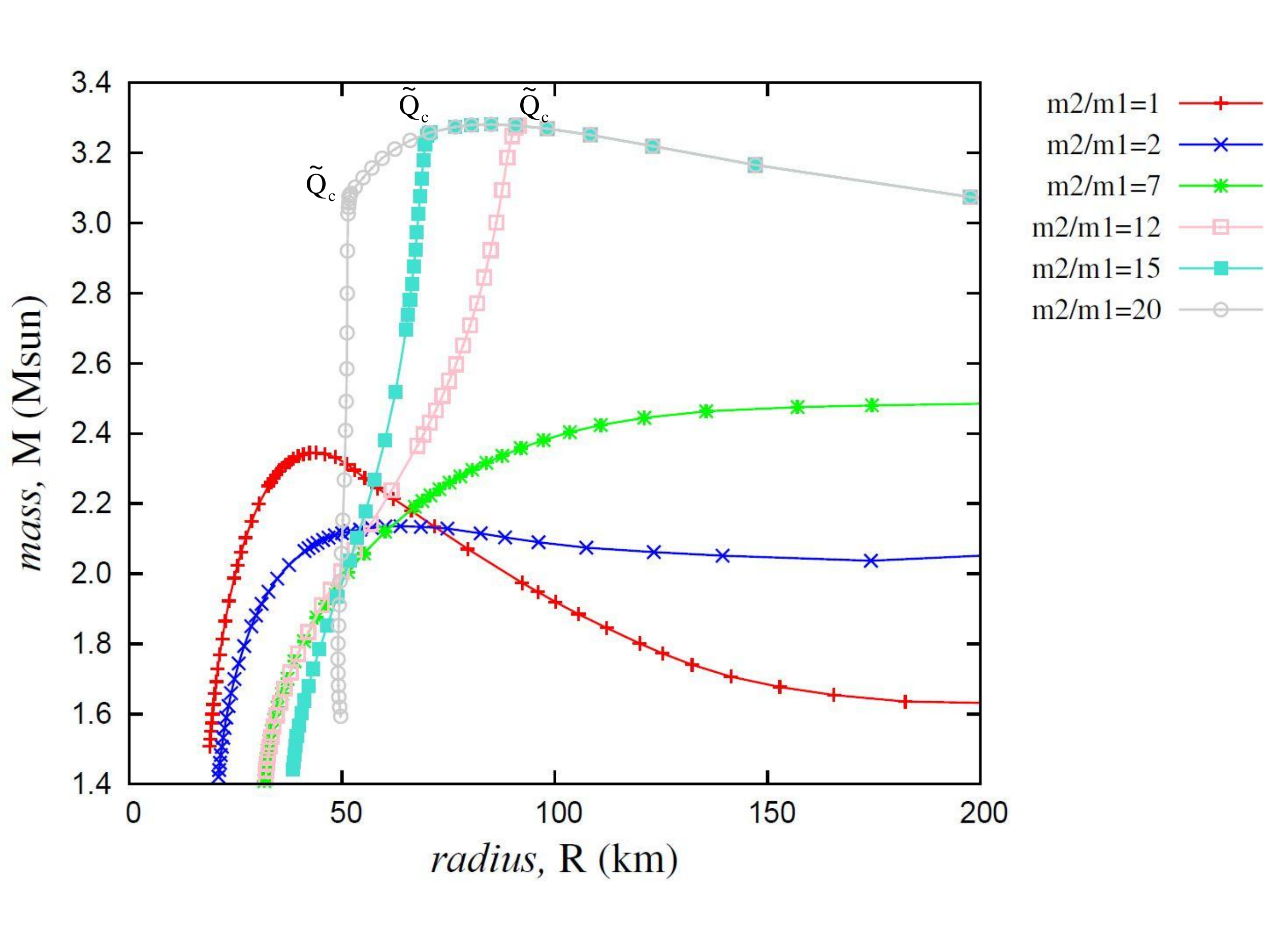}
\caption{Comparison of mass-radius relations for two flavor case with $\lambda=10$ and $M_\textrm{KK}=1.04$ GeV.
$\tilde{Q}_c$ denotes the transition density.}
\label{fig:MvRla10}
\end{center}
\end{figure}

We present the mass-radius relations obtained from the equations of state of these asymmetric mass ratio configurations, in Fig.~\ref{fig:MvRla10}.
The results show that there exist two distinct groups in the mass-radius relation curves depending on the value of the quark mass ratio.
Note that for small quark ratios, $m_2/m_1=1,2,7$, the matter consists of almost equal number of two quarks even at low densities, while  for $m_2/m_1=15,20$
there is a transition at a finite density from matter with $up$ quarks only  to matter with $up$ and $down$ quarks.
As well known,  matter with more flavors  has a softer EoS due to small Fermi pressure, leading to a compact star with lower maximum mass and smaller radius as shown in Fig.~\ref{fig:MvRla10}.
We can see the effect more clearly for the case with larger quark ratios. In most of the cases, at the transition point ($\tilde{Q}_c$), the mass of the holographic compact stars suddenly drops as we
increase the central density, i.e., as we approach the smaller radius.
As discussed in Sec.~\ref{sec:cond}, we may not have a well-defined Fermi sea since we are dealing with bosons. Nevertheless, it is interesting that our results
mimic the phenomena associated with the presence of a Fermi sea.

\begin{figure}
\begin{center}
\includegraphics[scale=0.5]{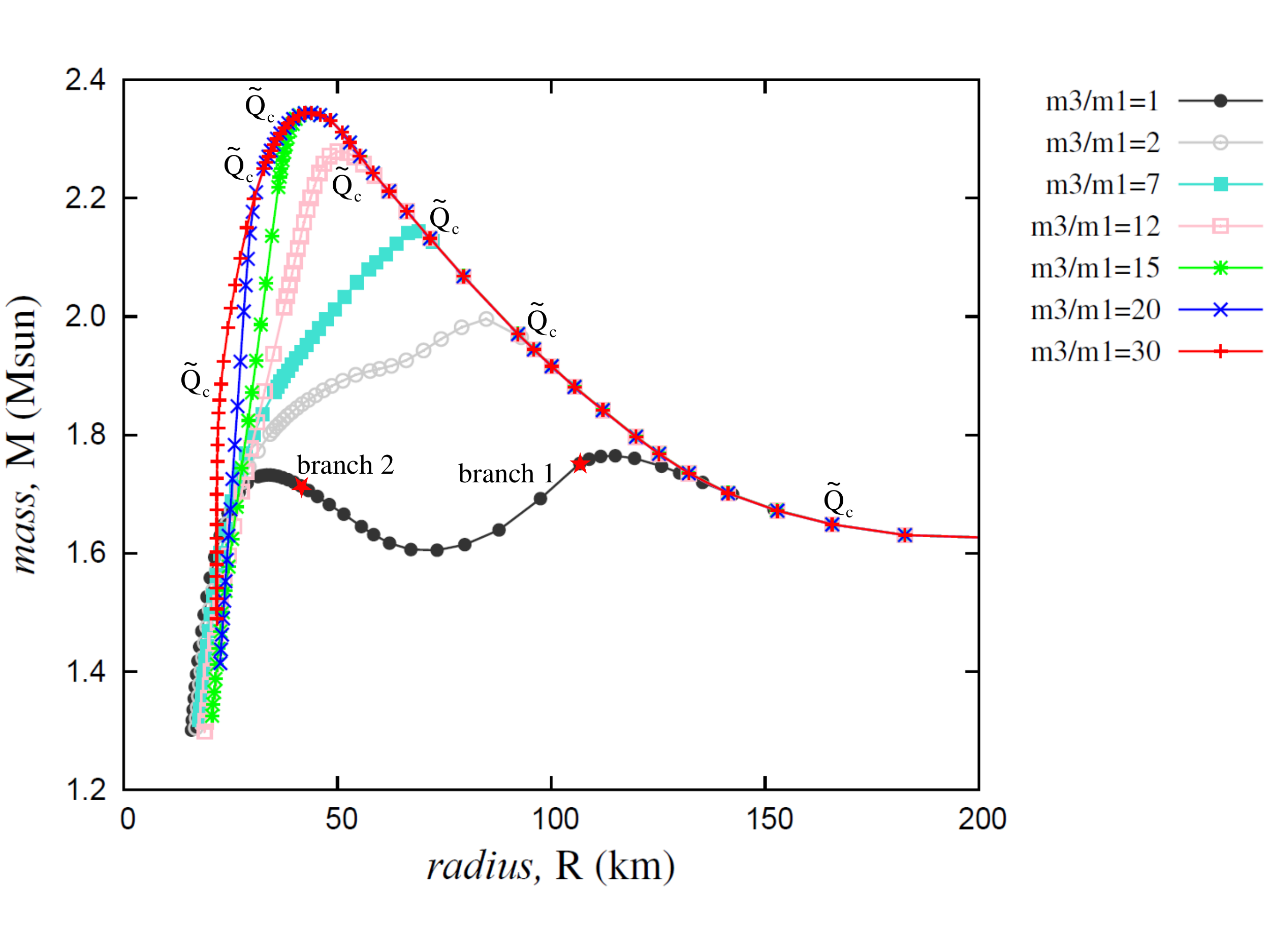}
\caption{Comparison of mass-radius relations for three flavor case with $\lambda=10$ and $M_\textrm{KK}=1.04$ GeV. ``branch $1$" denotes the first unstable branch of the EoS with $m_3/m_1=1$, whereas
``branch $2$" denotes the second stable branch. The red stars represent two star configurations discussed in this section.}
\label{fig:MvR21}
\end{center}
\end{figure}


Now we consider a three-flavor system  with masses $m_1$, $m_2$, and $m_3$.
Since the mass of the $strange$ quark is much larger than those of $up$ and $down$ quarks, we assume that
$m_1=m_2< m_3$.
We proceed as previous with two flavors, but here we fix $m_1=m_2=0.1$, which corresponds to $23.2$ MeV, and consider different mass ratios $m_3/m_1$.
Similar to the two-flavor case, there will be a transition from matter with only $up$ and $down$ quarks
to matter with $up$, $down$, and $strange$ quarks. In nature, this may correspond to nuclear matter to hyperon matter transition.
Note that we are in the confined phase since our background is the confining D4 geometry.
For illustration purpose, we choose $m_3/m_1=7,12,15,20,30$ and show the results in Fig.~\ref{fig:MvR21}.

\begin{figure}
\begin{center}
\includegraphics[width=10cm]{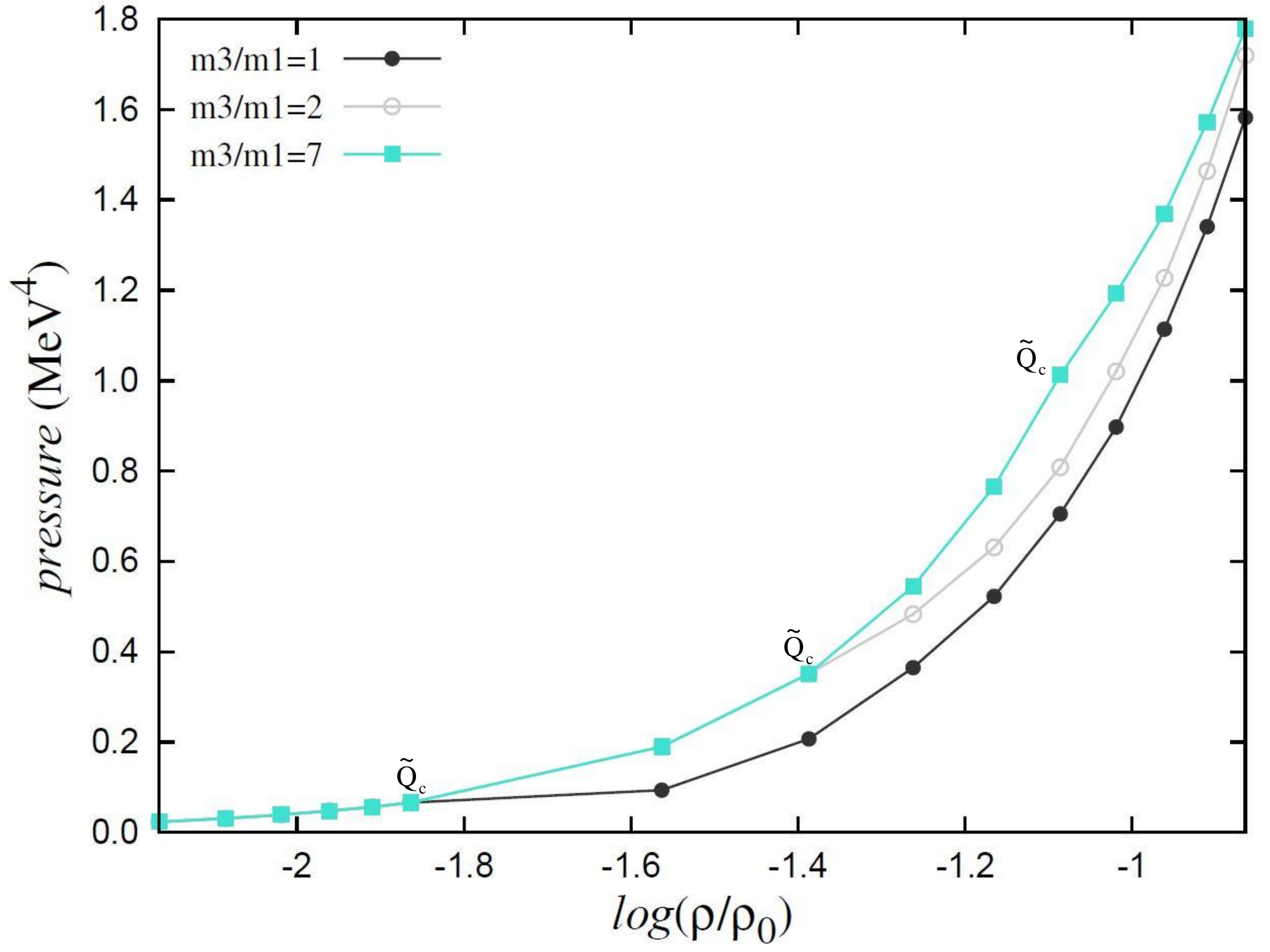}
\caption{The pressure versus matter density relations for the EoSs with $m_3/m_1=1,2$ and $7$.
In this region, the three EoSs transition from two-flavor to three-flavor matter, at different densities, denoted by $\tilde{Q}_c$ on the curves.}
\label{fig:PvRhodet}
\end{center}
\end{figure}

\begin{figure}
\begin{center}
\includegraphics[width=10cm]{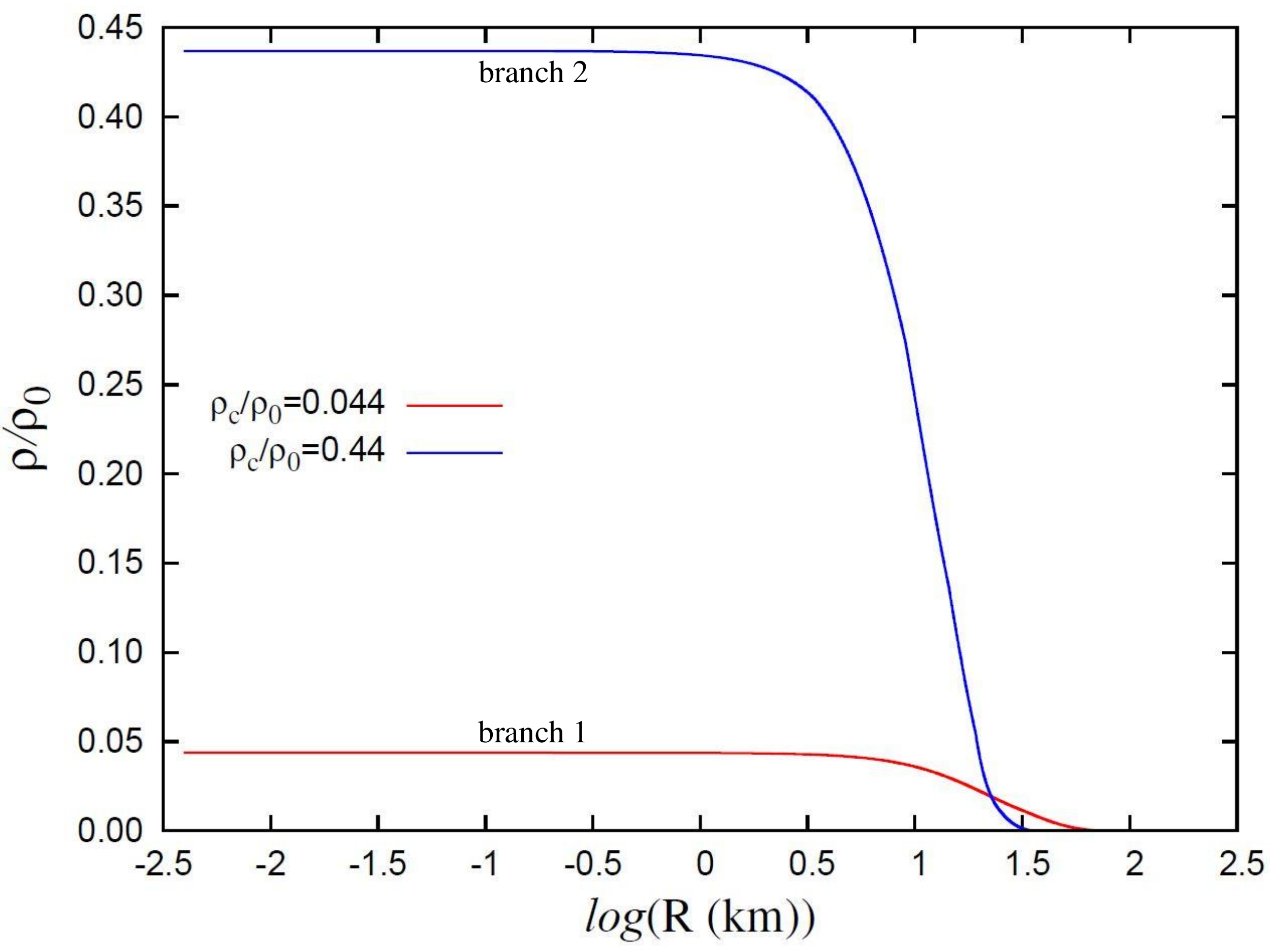}
\caption{The density profiles of the stars with the $m_3/m_1=1$ EoS at two central densities, one on the first unstable branch labeled $1$, and the other on the second stable branch labeled $2$.
The star configurations are represented by the red stars on Fig.~\ref{fig:MvR21}.}
\label{fig:rhoprof}
\end{center}
\end{figure}

As observed in the two-flavor case, there is a sudden change in the mass-radius curve at the transition density. This is due to the emergence of the third flavor in the ground state at the transition point.
For $m_3/m_1=1$, there is a dip in the mass-radius curve when the EoS undergoes the aforementioned transition, but at this mass ratio, the EoS
suddenly becomes softer when it transitions from two-flavor to three-flavor matter.
This is evident in the pressure versus matter density relation of the EoS in Fig.~\ref{fig:PvRhodet}, where the slope suddenly drops after the third flavor appears in the matter.
This region, where the EoS suddenly becomes softer after the transition, corresponds to the region in Fig.~\ref{fig:MvR21} where the $dM/dR$ is negative, i.e., branch $1$,
indicating that the star is unstable.
When the central density increases, the star becomes stabilized again, as the EoS becomes hard enough to support the piling up of matter.
We show the density profiles of two star configurations with this EoS in Fig.~\ref{fig:rhoprof}. The star on branch $1$ in Fig.~\ref{fig:MvR21} is much more dilute compared to
that on branch $2$, indicating that much more matter could be piled up in the latter compared to the former.

For $m_3/m_1=2$, however, this dip becomes much less pronounced, and as the mass ratio increases beyond this, the dip vanishes since at these mass ratios,
the EoS with only two flavors is comparable to that with three flavors right after the transition.
At these mass ratios, the slope of the pressure versus matter density relation does not drop as drastically as compared to the $m_3/m_1=1$ case after the matter transitions from two to three flavors.
Similar to the $m_3/m_1=1$ case, as the central density is increased further for these mass ratios, the three-flavor EoS becomes much harder, to the point where its stiffness exceeds that with two flavors.
The distinguishing point in these higher mass ratio cases is that the hardening occurs in a much shorter range of central densities after the transition, compared to the $m_3/m_1=1$ case.

As for the radii of the stars in Fig.~\ref{fig:MvR21} in general, we observe that they are still much larger than the realistic neutron star radii \cite{Lattimer:2013hma,Guver:2011qw,Guver:2011js}. However, since the masses are in the reasonable range, we believe that our work shows the applicability of holographic models to compact stars.

\section{Discussion}
\label{sec:sum}

In this work, we studied the effects of explicit flavor symmetry breaking due to quark mass differences on the holographic asymmetric dense matter and compact stars in the D4/D6 model.
Due to the fact that the D4/D6 model can easily incorporate non-zero quark masses, we   assume that the model may be suitable in investigating flavor physics, especially phenomena associated with explicit flavor symmetry breaking.
Previous works with the D4/D6 model in asymmetric dense matter \cite{Kim:2009ey, Kim:2011gw} seem to support this assumption.

We calculated the meson mass differences and the quark-antiquark condensates in asymmetric dense matter and found that our quark-antiquark condensates behave as if there exists a well-defined Fermi sea, at least near the transition density.
We investigated the infinite dense system to obtain the mass-radius relations of compact stars and observed that the masses of the holographic compact stars suddenly drop at the transition density as expected.

Even though our approach is very schematic, we showed the applicability of holographic QCD models in studying dense matter physics. In order to have more realistic holographic models, it will be important to have a clearer understanding of the relation between D-brane dynamics and results obtained in this work.
In addition, it will be interesting to see if one can describe the other physical quantities associated with explicit flavor symmetry breaking using holographic models.

\acknowledgments

The work of Y.Kim and I.J.Shin was supported by the Rare Isotope Science Project funded by the
Ministry of Science, ICT and Future Planning (MSIP) and National Research Foundation
(NRF) of KOREA (2013M7A1A1075766).
C.H.L. was supported by the BAERI Nuclear R \& D program (M20808740002) of MEST/KOSEF and the Financial Supporting Project of Long-term Overseas Dispatch of PNU's Tenure-track Faculty, 2013.

\bibliographystyle{apj}

\end{document}